# Teaching DevOps Security Education with Hands-on Labware: Automated Detection of Security Weakness in Python

Mst Shapna Akter[*], Juanjose Rodriguez-Cardenas[†], Md Mostafizur Rahman[†], Hossain Shahriar[‡], Akond Rahman[§], Fan Wu[¶]

[*]Dept. of Intelligence and Robotics Systems, University of West Florida, USA. Email: jannatul.shapna99@gmail.com
[†]Dept. of Information Technology, Kennesaw State University, USA. Email: jrodr225@students.kennesaw.edu, hshahria@kennesaw.edu
[†]Dept. of Information Technology, University of West Florida, USA. Email: md.mostafizur.rn@gmail.com
[‡]Center of Cyber Security, University of West Florida, USA. Email: hshahriar2012@gmail.com
[§]Dept. of Computer Science and Software Engineering, Auburn University, Auburn, USA. Email: akond@auburn.edu
[¶]Dept of Computer Science, Tuskegee University, Tuskegee, USA. Email: fwu@tuskegee.edu

*Abstract*—The field of DevOps security education necessitates innovative approaches to effectively address the ever-evolving challenges of cybersecurity. In adopting a student-centered approach, there is the need for the design and development of a comprehensive set of hands-on learning modules. In this paper, we introduce hands-on learning modules that enable learners to be familiar with identifying known security weaknesses, based on taint tracking to accurately pinpoint vulnerable code. To cultivate an engaging and motivating learning environment, our hands-on approach includes a pre-lab, hands-on and post lab sections. They all provide introduction to specific DevOps topics and software security problems at hand, followed by practicing with real world code examples having security issues to detect them using tools. The initial evaluation results from a number of courses across multiple schools show that the hands-on modules are enhancing the interests among students on software security and cybersecurity, while preparing them to address DevOps security vulnerabilities.

## I. INTRODUCTION

Hands-on learning is a widely accepted approach in education that revolves around learner-centered strategies and fosters active interaction among participants to enhance knowledge and analyzes case studies collaboratively [1]. This approach involves students engaging in discussions based on realistic scenarios that closely resemble real-world examples [2]. In order to equip students with essential skills through practical experiences in addressing real-world security challenges, hands-on learning employs a distinctive hands-on methodology that comprises pre-lab, lab, and post-lab activities [3]. Among these activities, the pre-lab stage introduces fundamental concepts related to the topic, enabling students to acquire the necessary background knowledge before undertaking practical work. The subsequent lab phase provides explicit instructions for hands-on practice, encouraging students to solve authentic problems using optimal solutions and fostering self-confidence through mastery-building experiences. The post-lab phase extends the hands-on practice, emphasizing the optimization and further development of solutions. This phase plays a crucial role in enhancing students' self-efficacy through observing peer performance and fostering creativity.

The proposed hands-on learning approach consists of the following steps (Figure 1 provides visual representations):
Step 1: Initiate/understand the topic through pre-lab instructions.

Step 2: Engage/analyze problems through hands-on lab activities involving real-life issues.

Step 3: Optimize the solutions obtained from the hands-on lab using various approaches.

Step 4: Repeat steps 1-3 with different algorithms or datasets.

Authentic learning is a teaching approach that immerses students in real-world tasks to develop practical skills and deeper understanding [4]. In the realm of software security, authentic learning has gained significant traction in recent times [5, 6]. For instance, Rahman et al. [7] utilized a learning platform to educate students on secure infrastructure-as-code development, while Lo et al. [8] pioneered authentic learning in the context of machine learning in cybersecurity, integrating portable hands-on labware.

Many institutions offer courses on cybersecurity in their curriculum. However, these courses often lack sufficient learning materials around DevOps security. DevOps is a software development approach that combines development (Dev) and operations (Ops) teams to streamline and automate the software delivery process [9]. Development of hands-on lab exercises pose several challenges, including a scarcity of knowledgeable instructors, complex configuration processes, the need for extensive resources and materials, and the commitment to completing all steps. In order to overcome these difficulties, we have developed an open-source, portable, modular, and easy-to-adopt hands-on learning modules for DevOps in the field of cybersecurity.

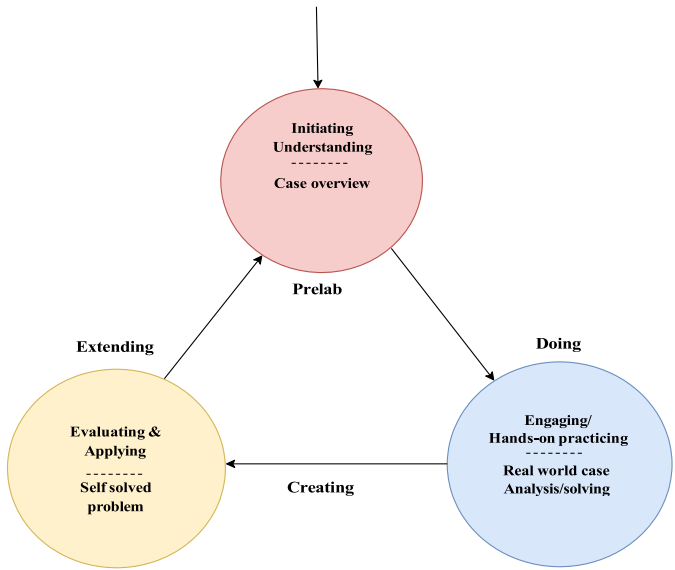

Fig. 1: Steps of Hands-on Learning Approach

Our approach is readily available online and is structured into 10 learning modules, each consisting of three parts: Pre-lab, hands-on lab, and post-lab activities. These modules cover a wide range of topics related to DevOps and cybersecurity, including an Installation Overview (M0), Automated Requirements Validation (M1), Automated Detection of Known Security Weaknesses (M2), Automated Taint Tracking for Accurate Detection (M3), Automated Forensicability (M4), Git Hooks to Facilitate Automated Security Static Analysis (M5), Security Weakness Identification with Continuous Integration (M6), Security Weaknesses in Infrastructure as Code Scripts (M7), Security Weaknesses in Kubernetes Manifests (M8), Chaos Engineering with White-box Fuzzing (M9), and Automated Secret Management (M10). These learning modules provide students with comprehensive resources and practical exercises to enhance their knowledge and skills in applying DevOps principles to cybersecurity. By utilizing our open-source approach, students can gain hands-on experience and develop proficiency in addressing real-world security challenges within the DevOps framework.

A security vulnerability is a type of bug that can compromise computer systems, programs, or mobile and web applications [10]. DevOps Security vulnerabilities can take various forms, including design flaws, programming errors, or incorrect configurations, and they can result in unauthorized access, data breaches, or service disruptions. If a security vulnerability exists in an application or system, it can be exploited by malicious actors to gain unauthorized access or manipulate the system's behavior without the user's knowledge or consent [11]. These vulnerabilities may arise from various sources, such as insecure coding practices, weak authentication mechanisms, or inadequate input validation. Therefore, identifying and addressing security vulnerabilities is crucial to ensuring the overall security and resilience of computer systems and applications.

Taint tracking is an effective approach in uncovering bugs or vulnerabilities with greater precision. Taint tracking involves tracing the flow of sensitive or untrusted data throughout a system or application. By marking or "tainting" specific data inputs as they enter the system, developers can track how this tainted data propagates and interacts with different components. This allows for the identification of potential security flaws or vulnerabilities that could be exploited by malicious actors. Taint tracking provides valuable insights into how data is processed, manipulated, and potentially misused within a system, enabling developers to pinpoint and address vulnerabilities more accurately. By incorporating taint tracking techniques into the security analysis process, students can learn how to detect and remediate bugs or vulnerabilities effectively. Taint tracking provides a critical layer of analysis that goes beyond traditional methods, ensuring a more thorough and accurate assessment of potential security risks. In this paper, we introduce hands-on learning modules that enables learners to be familiar with identifying known security weaknesses, based on taint tracking to accurately pinpoint vulnerable code. We introduce various steps for automated detection of security weakness in python code. The initial evaluation results from a number of courses across multiple schools show that the hands-on modules are enhancing the interests among students on software security and cybersecurity, while preparing them to address DevOps security vulnerabilities.

This paper is organized as follows. Section II provides an overview of the related work in the field. Section III presents the design of the labware, divided into three parts: Pre-lab, Hands-on lab, and Post-lab. The student learning assessment is discussed in Section IV. Finally, Section V concludes the paper and summarizes the key findings and contributions.

## II. RELATED WORK

From the literature review we have found that several studies that have implemented case study-based [12], project-based, and authentic learning approaches in various disciplines, including cybersecurity and software engineering. For instance, Deng et al. [13] implemented a case-study-based learning approach for machine learning-based hands-on-lab exercises in cybersecurity. Similarly, Blanken et al. [14] performed a case-study-based module to engage learners in ethical dilemmas in cybersecurity, while Garg et al. [15] compared case study-based and lecture-based approaches in software engineering research. Frontera et al. [16] followed a project-based learning framework to evaluate cyber-attacks on a cyber-physical system, and Huang et al. [17] integrated applied machine learning technology with project-based learning in engineering programs.

In addition, Lo et al. [6] conducted an authentic learning project in the field of cybersecurity, involving students in solving real-world cybersecurity challenges. Faruk et al. [18]

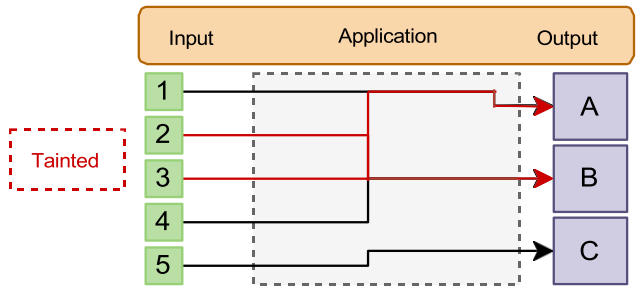

Fig. 2: Taint Tracking System

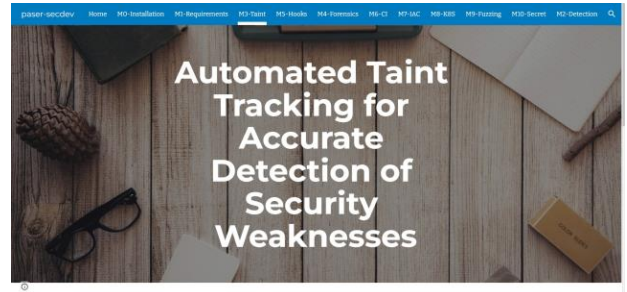

Fig. 3: Screenshot of module 3

focused on the implementation of authentic learning in ML for cybersecurity, developing learning modules with hands-on activities for solving security problems. Qian et al. [5] implemented an authentic learning approach in secure software development, providing students with hands-on laboratory practice in secure mobile app development.

Despite the valuable contributions of these studies, it is worth noting that none of them specifically addressed DevOps security education. While case study-based, project-based, and authentic learning approaches have been implemented in various disciplines, there is a gap in the literature regarding the application of these approaches in the context of DevOps security education. Therefore, our work fills this gap by developing authentic learning modules specifically tailored for DevOps security education.

## III. LABWARE DESIGN

The portable labware has been developed, designed, and deployed on the open-source environment GitHub. This environment allows users to access and share resources from anywhere and at any time without installation and maintenance hassles. As a result, learners can collaboratively interact with their peers, practice, and run all modules. Each of our hands-on labware modules is developed using vulnerable codes. The structure of these modules comprises three components: a pre-lab section covering basic knowledge, a hands-on lab section providing in-depth explanations of experiments, and a post-lab section offering instructions for further optimizations.

### A. Pre-lab

The Pre-Lab outlines DevOps solutions to software security, including prevention and detection, and introduces a specific software security study scenario with the root causes of security threats, attack plans, and their effects. Students can gain perspective and insight by observing a simplified "hello world" example for the software security case and its corresponding DevOps solution. This prepares students with a specific softwaresecurity case for conceptual understanding and provides them with a starting point to experience DevOps solutions for such software security cases. By utilizing the DevOps technique, it helps students develop a fundamental understanding of why these software security vulnerabilities need to be addressed. Figure 3 shows a screenshot of the Pre-Lab for Module 2's Automated Detection of Known Security Weaknesses.

### B. Hands-on lab

The open-source GitHub platform, which is an in-browser environment accompanied by a free Google Cloud service, is utilized for designing, creating, and deploying the hands-on activity laboratories. Upon completion of the practical activity lab, students will gain valuable first hand experience in problem-solving. Students can further enhance their understanding by utilizing visual cues and screenshots provided for each stage.

### C. Post-lab

Students are encouraged to use their newly acquired knowledge and skills to address actual problems in the Post add-on lab. It encourages critical reflection on the provided example and practical application for improving problem-solving, such as raising the prediction and detection accuracy rate with new innovative concepts and active testing and experiments. With Colab, students can share their original work with others on the cloud. Colab, short for Google Colaboratory, is an online platform that provides free access to a Jupyter notebook environment along with GPU support, enabling users to write, share, and run code collaboratively [19].

## IV. STUDENT LEARNING ASSESSMENT

We implemented the module in three schools during spring2023. A preliminary survey collected from a total of seventy two undergraduate Engineering students at Kennesaw State University, Auburn University, Tuskegee University. Surveys are represented in quantitative and qualitative views. We conducted both a prelab and postlab survey, where we asked various questions.

**Prelab Survey:** Among the seventy four students surveyed, the majority (55) considered themselves to be in the age group between 18 and 25 years. A few of them (16) fell into the age group between 26 and 35, two of them between 36 and 45, and one of them between 46 and 55. We asked the participants to describe their level of education in the field of DevOps Security in Figure 4. Additionally, we inquired about their preferences regarding (a) project-based lab work versus listening to lectures, (b) personally doing or working through examples, and (c) having a learning/tutorial system that provides feedback. The responses are displayed in Figures 4 and 5. In Fig 4, we posed four questions: (a) Which course are you enrolled in? (b) What is your gender? (c) What is your race? and (d) Do real-world relevant applications engage your learning in cybersecurity? For question (a), 25 participants were enrolled in Programming I and II at Tuskegee University, 16 in Security Concepts at KSU, 13 in Information Security at Tuskegee University, 10 in IT 4823 - Information Security Concepts, 7 in Physical IT System Security at KSU, 2 in Data Analytics, and 2 in IT 6413 at KSU. All the participants come from a STEM (Science, Technology, Engineering, and Mathematics) background, indicating they have a foundational knowledge of technology. For question (b), 44 participants responded as male, while 30 participants responded as female. For question (c), 44 participants identified themselves as African American, 21 as Asian, and 5 as white. For question (d), 34 participants agreed, 12 strongly agreed, and 12 were neutral. In Fig 5, three questions are presented: (a) "I learn better by listening to lectures." (b) "Please indicate the extent to which you have received education in the following areas based on the given scale: i. DevOps Security or IaC security, ii. Software Engineering, iii. Software Cybersecurity." (c) "For each of the following statements, indicate the extent to which you agree or disagree: i. I learn better by engaging in hands-on lab work, ii. I learn better by listening to lectures, iii. I learn better by personally doing or working through examples, iv. I learn better by reading the material on my own, v. I learn better by having a learning/tutorial system that provides feedback." For question (a), most respondents indicated they had no experience in programming languages (35 to 40 participants). A few had limited or moderate experience, 5 to 10 had good experience, and none identified as expert programmers. For question (b), most participants reported having no education in DevOps security, while only a few had experience in Software Engineering and Software Cybersecurity. For question (c), the majority of participants strongly agreed with statements i, iii, and v.

TABLE I: Display the responses of students on age group

| # | Answer | % | Count |
|---|---|---|---|
| 1 | < 18 years | 0.00% | 0 |
| 2 | Between 18 and 25 years | 74.32% | 55 |
| 3 | Between 26 and 35 years | 21.62% | 16 |
| 4 | Between 36 and 45 years | 2.70% | 2 |
| 5 | Between 46 and 55 years | 1.35% | 1 |
| 6 | >55 years | 0.00% | 0 |

**Post-Test Survey**: We asked students if the tutorials in the pre-lab helped them understand more about the topics; in a post-test survey we completed it after involvement in the practical lab. Figure 6 and 7 displays the responses. In Figure 6, feedback on the secure DevOps materials revealed a predominantly positive response. For the statement, "I like being able to work with the secure DevOps hands-on materials," 28 participants agreed, 14 strongly agreed, 18 were neutral, 1 disagreed, and 1 strongly disagreed. For the statement, "The tutorials help me learn more on the topic," 34 agreed, 10 strongly agreed, 16 remained neutral, and 5 disagreed. For the statement, "hands-on labs help in understanding DevOps security better," 32 agreed, 12 strongly agreed, 16 were neutral, and 5 disagreed. Lastly, for the statement, "The hands-on labs enhance my learning on secure DevOps coding and best practices," 32 participants agreed, 14 strongly agreed, 16 were neutral, and 2 disagreed.

In Figure 7, the feedback on the secure DevOps materials indicated an overall positive sentiment. For the statement "The real-world relevant applications engage my learning on cybersecurity," 34 participants agreed, 14 strongly agreed, 14 were neutral, and 1 strongly disagreed. Regarding the assertion, "The learning modules help me apply learned knowledge to solve cybersecurity problems in the future," 32 participants agreed, 10 strongly agreed, 15 were neutral, 2 disagreed, and 2 strongly disagreed. In response to "The post-lab motivates and promotes me to continue studying," 29 participants agreed, 10 participants strongly agreed, 17 were neutral, 5 disagreed, and 2 strongly disagreed.

The feedback gathered from students post their engagement with the secure DevOps hands-on materials and real-world relevant applications showcases a notably affirmative inclination towards the efficacy and impact of these materials. As seen in Figure 6, a significant majority of the participants acknowledged the value of the secure DevOps materials, particularly in aiding their understanding of the subject matter. A standout observation is that more than 80% of the participants either agreed or strongly agreed that the hands-on labs and tutorials were beneficial to their learning, especially in the realm of secure DevOps coding best practices and understanding DevOps security.

In Figure 7, the sentiment continues to be positive, underscoring the relevance and effectiveness of the provided materials. A substantial 48 out of 74 participants (over 70%) agreed or strongly agreed that real-world applications enhanced their cybersecurity learning. Similarly, a combined total of 42 participants confirmed the efficacy of the learning modules in applying their acquired knowledge to future cybersecurity challenges. The post-lab feedback also suggests that the lab experience serves as a motivational tool, with a majority expressing that it encourages them to delve deeper into their studies.

These findings emphasize the crucial role of practical hands-on materials and real-world applications in fostering an enriched and engaged learning experience for students, enhancing not only their comprehension but also their enthusiasm to continue their studies in the field of cybersecurity and DevOps.

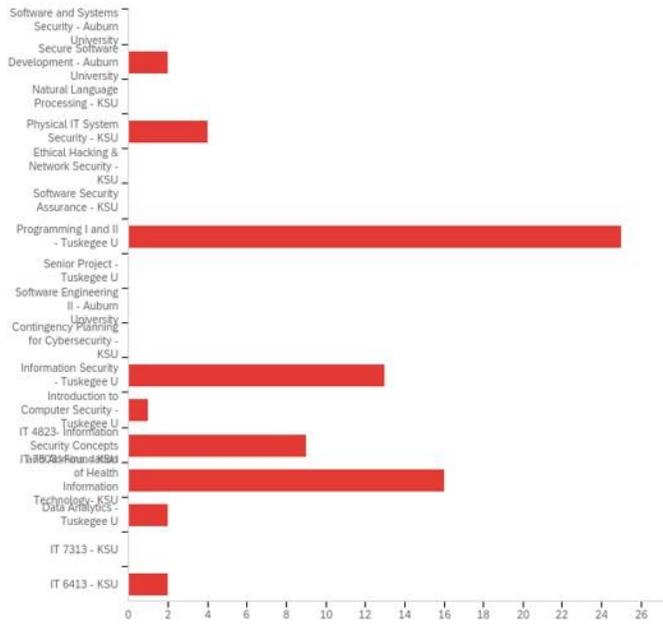
(a)

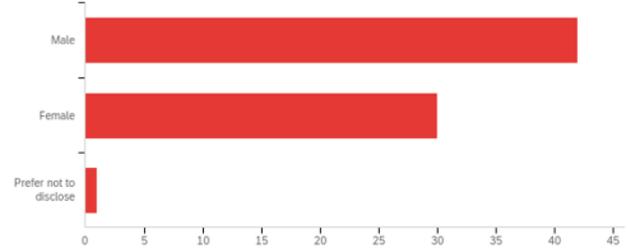
(b)

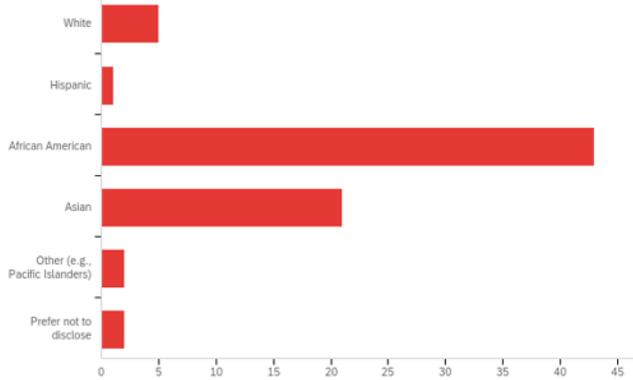
(c)

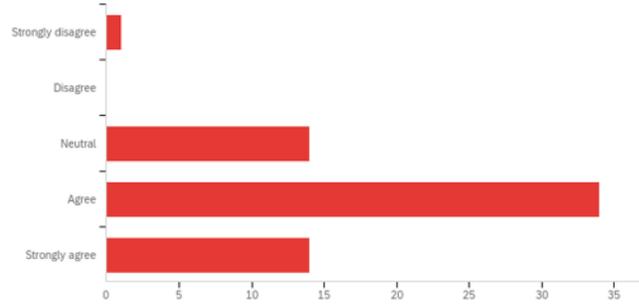
(d)

Fig. 4: Figure (a), (b), (c) and (d) displays the reponses from the pre-survey questions.

## V. CONCLUSION

This labware aims to address the challenges and requirements of learning DevOps for security by utilizing effective and engaging authentic learning techniques, as well as filling the gap in pedagogical resources and hands-on learning environments. The project introduces a novel teaching approach that utilizes DevOps to proactively resolve security issues. Based on preliminary feedback, students not only grasp the concepts but also practice the skills through the hands-on laboratories.

## ACKNOWLEDGEMENT

The work is supported by the National Science Foundation under NSF Award #2100134, #2100115, #2209638, #2209637, #1663350, #2310179. Any opinions, findings, recommendations, expressed in this material are those of the authors and do not necessarily reflect the views of the National Science Foundation.

## REFERENCES

[1] B. Z. Harvey, R. T. Sirna, and M. B. Houlihan, "Learning by design: Hands-on learning.," *American School Board Journal*, vol. 186, no. 2, pp. 22–25, 1998.

[2] C. F. Herreid, *Start with a story: The case study method of teaching college science*. NSTA press, 2007.

[3] M. S. Akter, H. Shahriar, S. I. Ahamed, K. D. Gupta, M. Rahman, A. Mohamed, M. Rahman, A. Rahman, and F. Wu, "Case study-based approach of quantum machine learning in cybersecurity: Quantum support vector ma-

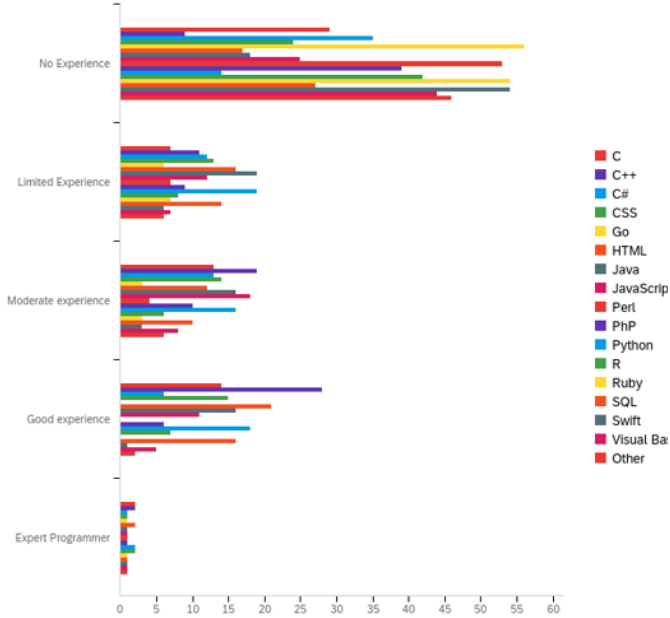

(a)

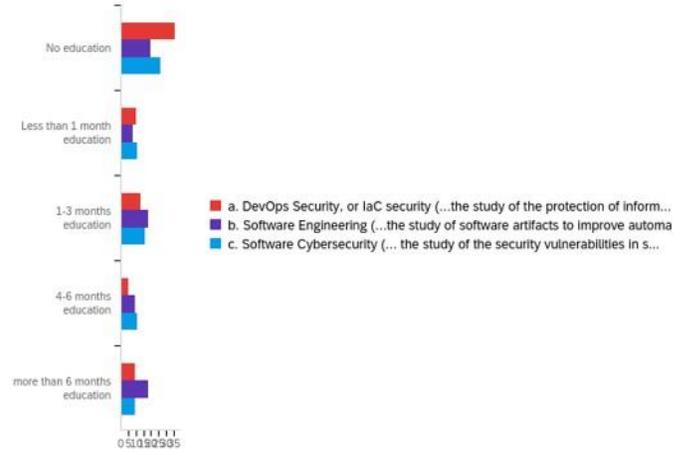

(b)

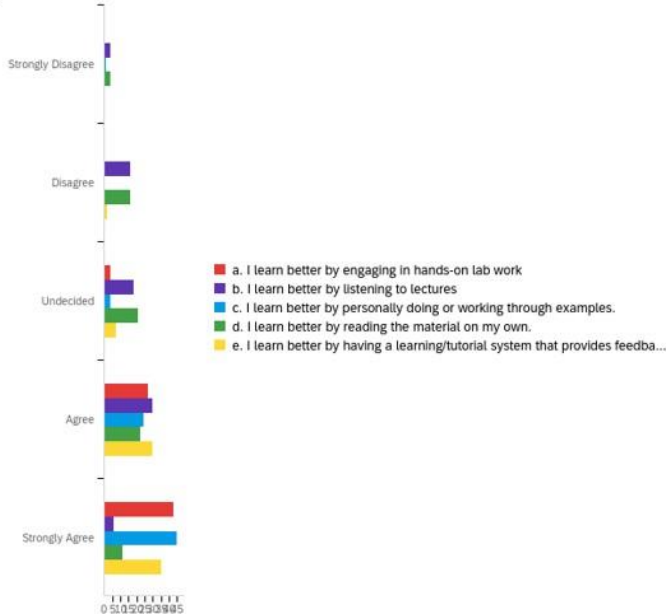

(c)

Fig. 5: Figure (a), (b), and (c) displays the reponses from the pre-survey questions.

chine for malware classification and protection," *arXiv preprint arXiv:2306.00284*, 2023.

[4] J. Herrington and R. Oliver, "An instructional design framework for authentic learning environments," *Educational technology research and development*, vol. 48, no. 3, pp. 23–48, 2000.

[5] K. Qian, D. Lo, R. Parizi, F. Wu, E. Agu, and B.-T. Chu, "Authentic learning secure software development (ssd) in computing education," in *2018 IEEE Frontiers in Education Conference (FIE)*, pp. 1–9, IEEE, 2018.

[6] D. C.-T. Lo, K. Qian, W. Chen, H. Shahriar, and V. Clincy, "Authentic learning in network and security with portable labs," in *2014 IEEE Frontiers in Education Conference (FIE) Proceedings*, pp. 1–5, IEEE, 2014.

[7] A. Rahman, S. I. Shamim, H. Shahriar, and F. Wu, "Can we use authentic learning to educate students about

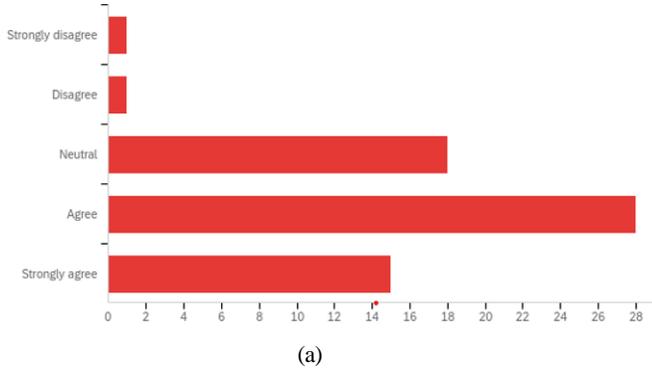
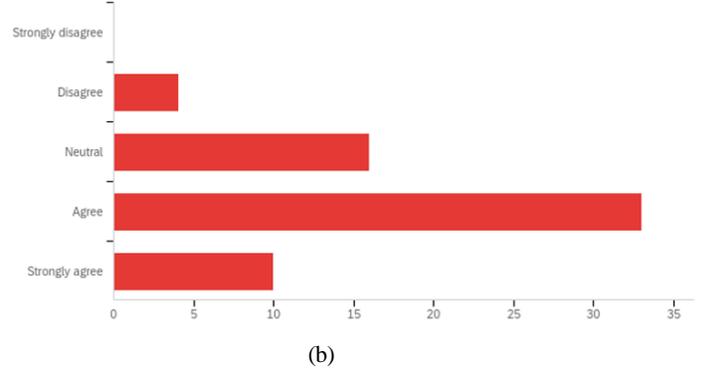
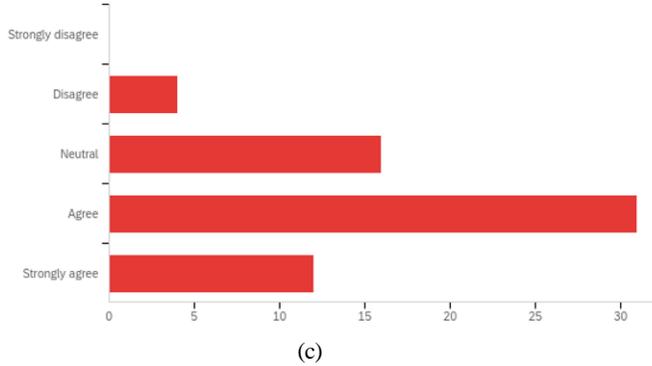
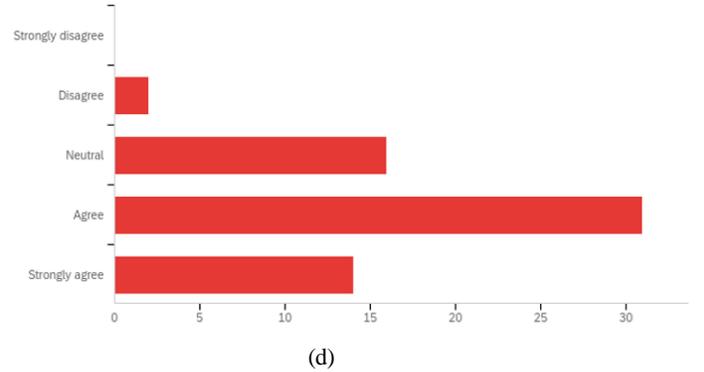

Fig. 6: Figure (a), (b), (c), and (d) displays the responses from the post-survey questions.

secure infrastructure as code development?," in *Proceedings of the 27th ACM Conference on on Innovation and Technology in Computer Science Education Vol. 2*, pp. 631–631, 2022.

[8] D. C.-T. Lo, H. Shahriar, K. Qian, M. Whitman, F. Wu, and C. Thomas, "Authentic learning of machine learning in cybersecurity with portable hands-on labware," in *Proceedings of the 53rd ACM Technical Symposium on Computer Science Education V. 2*, pp. 1153–1153, 2022.

[9] C. Ebert, G. Gallardo, J. Hernantes, and N. Serrano, "Devops," *Ieee Software*, vol. 33, no. 3, pp. 94–100, 2016.

[10] M. S. Akter, M. J. H. Faruk, N. Anjum, M. Masum, H. Shahriar, N. Sakib, A. Rahman, F. Wu, and A. Cuzzocrea, "Software supply chain vulnerabilities detection in source code: Performance comparison between traditional and quantum machine learning algorithms," in *2022 IEEE International Conference on Big Data (Big Data)*, pp. 5639–5645, IEEE, 2022.

[11] M. S. Akter, H. Shahriar, and Z. A. Bhuiya, "Automated vulnerability detection in source code using quantum natural language processing," in *Ubiquitous Security: Second International Conference, UbiSec 2022, Zhangjiajie, China, December 28–31, 2022, Revised Selected Papers*, pp. 83–102, Springer, 2023.

[12] M. Shapna Akter, H. Shahriar, S. I. Ahamed, K. Datta Gupta, M. Rahman, A. Mohamed, M. Rahman, A. Rahman, and F. Wu, "Case study-based approach of quantum machine learning in cybersecurity: Quantum support vector machine for malware classification and protection," *arXiv e-prints*, pp. arXiv–2306, 2023.

[13] Y. Deng, D. Lu, D. Huang, C.-J. Chung, and F. Lin, "Knowledge graph based learning guidance for cybersecurity hands-on labs," in *Proceedings of the ACM conference on global computing education*, pp. 194–200, 2019.

[14] J. Blanken-Webb, I. Palmer, S.-E. Deshaies, N. C. Burbules, R. H. Campbell, and M. Bashir, "A case study-based cybersecurity ethics curriculum," in *2018 USENIX Workshop on Advances in Security Education (ASE 18)*, 2018.

[15] K. Garg and V. Varma, "A study of the effectiveness of case study approach in software engineering education," in *20th Conference on Software Engineering Education & Training (CSEET'07)*, pp. 309–316, IEEE, 2007.

[16] P. J. Frontera and E. J. Rodríguez-Seda, "Network attacks on cyber–physical systems project-based learning activity," *IEEE Transactions on Education*, vol. 64, no. 2, pp. 110–116, 2020.

[17] L. Huang, "Integrating machine learning to undergraduate engineering curricula through project-based learning," in *2019 IEEE Frontiers in Education Conference (FIE)*,

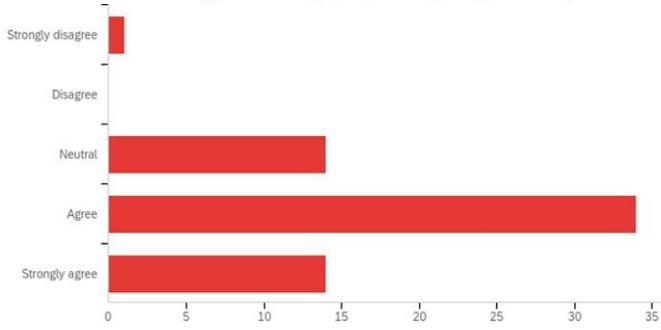
(a)
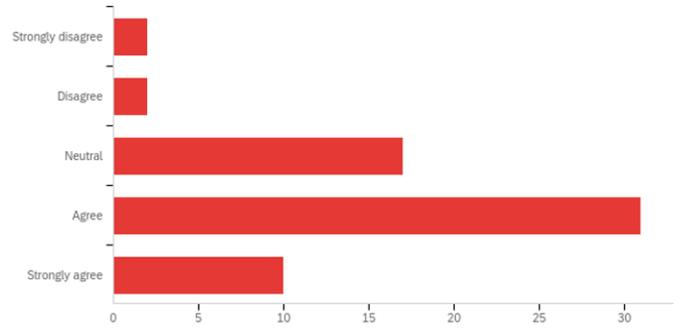
(b)
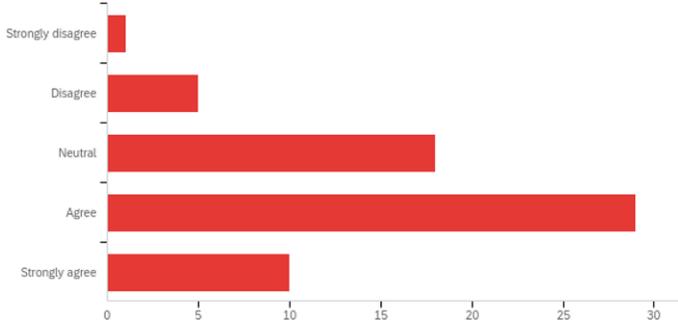
(c)

Fig. 7: Figure (a), (b), and (c) displays the responses from the post-survey questions.

pp. 1–4, IEEE, 2019.

[18] M. J. H. Faruk, M. Masum, H. Shahriar, K. Qian, and D. Lo, "Authentic learning of machine learning to ransomware detection and prevention," in *2022 IEEE 46th Annual Computers, Software, and Applications Conference (COMPSAC)*, pp. 442–443, IEEE, 2022.

[19] E. Bisong and E. Bisong, "Google colaboratory," *Building machine learning and deep learning models on google cloud platform: a comprehensive guide for beginners*, pp. 59–64, 2019.